\documentclass[nolineno]{gd-llncs-short}
\usepackage{graphicx}
\usepackage{float}
\usepackage[title]{appendix}
\usepackage{color,soul}
\usepackage{subcaption}

\begin{document}

\title{Can an NN Model plainly learn Planar Layouts?}

\author{Simon van Wageningen\orcidID{0000-0002-0346-5597}\and\\
Tamara Mchedlidze\orcidID{0000-0001-6249-3419}}
%
%
\institute{Utrecht University, Utrecht, The Netherlands\\ \email{\{s.vanwageningen, t.mtsentlintze\}@uu.nl}}

\maketitle              
%


\subsubsection{Introduction}

Planar graph drawings tend to be aesthetically pleasing~\cite{ref_qualitymetric_interpretability}. Planar graphs and their drawings have been extensively studied in graph drawing literature~\cite{planarity_test} and can be generated efficiently~\cite{planarity_test,handbook_planar}. However, there are no practical layout algorithms for the graphs that are nearly planar. Thus, force directed algorithms~\cite{force_directed} often fail at detecting the planar substructure in such graphs. The attempts to formalize near-planarity (1-planar~\cite{1_planar}, RAC~\cite{RAC} and quasi-planar~\cite{quasi_planar}) lead to NP-hard recognition problems~\cite{NP_hard,NP_complete}.

Due to the fact that the formalization of near-planarity immediately leads to NP-hard problems, we turn our attention towards Neural Networks (NNs). NNs have already been used for graph layout evaluation~\cite{eval1,eval2}, discrimination~\cite{discrim} and more relevantly for graph layout generation~\cite{kwon_ma,giovann,deepdrawing}.

Our far-reaching goal is to investigate whether NNs are capable of producing drawings of nearly planar graphs that clearly depict large planar substructures. Such NNs are expected to be able to produce near-planar drawings of planar graphs. Therefore, as a first step towards our goal, we investigate whether NNs are successful in producing planar drawings of planar graphs. Additionally, we briefly explore the effectiveness of the model in generalizing beyond planarity.

\subsubsection{Method}
We refer the reader to the appendices for the details of the experiments.
We reuse the LSTM model and Procrustes Statistic\footnote{$PS$ LF calculates the differences between original and predicted node coordinates, after a series of transformations} ($PS$) loss function (LF) of Wang et al.~\cite{deepdrawing,deepdrawing_github}, who showed the model to be successful in producing planar drawings of grids and stars. Note that the PS LF ensures that coordinate-based patterns can be learned.

Since drawings with less stress are shown to correlate with positive preferences~\cite{userstudy_stress}, additional experiments are also conducted using a supervised stress\footnote{Differences between original and predicted pairwise stress values, for each node pair, are computed} ($SuS$) LF. We expect the $SuS$ LF to be more capable than the $PS$ LF when randomness is introduced to node coordinates.

We train 8 models on 8 different graph classes and layouts: Grids, Grids with all diagonals ($Grids_d$), Grids with random diagonals ($Grids_{rd}$), Delaunay Triangulations, 2-star caterpillar ($Caterp2$), 3-star caterpillar ($Caterp3$), randomized radial trees ($RR Trees$) and randomized Stress Majorization trees ($RSM Trees$). The number of graphs in a dataset (72-1000) and the graph sizes (18-625 nodes) vary, depending on the graph class. Moreover, the testing datasets are comprised of multiple instances of similar sized graphs, as to make valid averaged-out comparisons. We evaluate the performance by visually inspecting the layouts and computing three quality metrics: the number of crossings ($nc$), the stress ($s$) and the angular resolution~\cite{angular_resolution} ($ar$). The quality metrics values are compared with two conventional layout techniques: ForceAtlas2~\cite{forceatlas2,forceatlas2_python} ($FD$) and Stress Majorization~\cite{stress_majorization} ($SM$).


\subsubsection{Results}


\begin{table}[t!]
\caption{Averaged performance of conventional techniques and model with $SuS$ LF on multiple instances of different graph classes. Stress $s$ is in 1\textbf{e+7}, bolded entries indicate interesting differences.\label{tab1}}
\begin{tabular}{||l|l||l|l|l|l||l|l|l||l|l|l||}
\hline
 &  &  &  &  &  & FD &  &  & SM &  & \\
\hline
Graph class & & LF & QM & QM & QM &  QM & QM & QM &  QM & QM & QM\\
\hline
Train & Test & SuS & nc & s & ar &  nc & s & ar & nc & s & ar\\
\hline
\hline
Grids & Grids &  1.54 & \textbf{6.30} & 4.62 & 0.36 & \textbf{15.90} & 6.18 & 0.50 & \textbf{0.61} & 4.88 & 0.93\\
Grids$_{d}$ & Grids$_{d}$ & 3.27 & \textbf{303} & 9.76 & 0.23 & \textbf{417} & 12.40 & 0.19 & \textbf{426} & 9.55 & 0.73\\
Grids$_{rd}$ & Grids$_{rd}$ & 2.60 & 3.31 & 4.24 & 0.21 & 27.5 & 6.10 & 0.26 & 3.04 & 4.82 & 0.57\\
Grids & Grids$_{rd}$ & 12.80 & 194 & 4.65 & \textbf{0.0028} & 27.5 & 6.10 & \textbf{0.26} & 3.04 & 4.82 & \textbf{0.57}\\
Grids$_{d}$ & Grids$_{rd}$ & 55.40 & 489 & 3.02 & 0.0046 & 27.5 & 6.10 & 0.26 & 3.04 & 4.82 & 0.57\\
\hline
Delaunay & Delaunay & 21.80 & 200 & 3.25 & 0.0086 & 59.40 & 3.57 & 0.026 & 90 & 3.10 & 0.038\\
\hline
Caterp2 & Caterp2 & 39.80 & 0 & 4.64\ & 0.051 & 0 & 4.24 & 0.090 & 0 & 5.39 & 0.15\\
Caterp3 & Caterp3 & 34.50 & 0.14 & 3.48 & \textbf{0.089} & 0.39 & 3.75 & \textbf{0.061} & 0 & 3.98 & \textbf{0.067}\\
\hline
RR Trees & RR Trees & 50.00 & 51.30 & 6.41 & 0.0054 & 3.87 & 5.13 & 0.052 & 30.90 & 5.63 & 0.13\\
RSM Trees & RSM Trees & 27.30 & 55.90 & 5.63 & 0.0036 & 3.68 & 5.28 & 0.048 & 32.80 & 5.66 & 0.13\\
\hline
\end{tabular}
\end{table}

Table~\ref{tab1} showcases the results of the experiments with the $SuS$ LF. On average, the models trained with the $SuS$ LF outperform the models trained with the $PS$ LF. Additionally, the model trained on Grids with the $SuS$ LF outperforms the $FD$ algorithm, in terms of number of crossings ($nc$) and stress ($s$). The model trained on $Caterp3$ shows a better angular resolution ($ar$) and stress than the $FD$ and $SM$ layouts. On average, the models trained with $SuS$ show better stress scores than the conventional $FD$ and $SM$ techniques. However, w.r.t. the $ar$ and the $nc$ the results tend to worsen. Moreover, when some randomness is introduced to the training data ($RRTrees$ \& $RSMTrees$), the models have difficulties generalizing, produce sub-optimal layouts and have unfavorable $QM$ results. When it comes to generalizing beyond planarity, a model trained on Grids and tested on Grids$_{rd}$ shows poor results. 

To conclude, our results indicate that planar graph classes can be learned by a Neural Network, and the produced planar drawings can score better than those produced by conventional techniques. We note that the loss function and the presence of randomness in graph data can have major effects on the model's learning capabilities. In the future, the combination of multiple loss functions should be explored as well as different Neural Network architectures.


\begin{subappendices}
\renewcommand{\thesection}{\Alph{section}}%

\section{LSTM}
The model used by Wang et al.~\cite{deepdrawing} is a Long Short-Term Memory (LSTM) model, which is a variant of Recurrent Neural Networks (RNNs). Through the use of various gates (input, output and forget) the LSTM model is capable of learning long-distance dependencies in sequential data. 

In order to have a graph as input for such a model, the graph data has to be transformed to a sequence. This transformation can be done using Breadth-First Search (BFS) starting from a random node in the graph. The resulting sequence is then given as input in the model. At each step of the input sequence, the LSTM model is supplied with additional graph information in the form of node feature vectors. These feature vectors encode the connectivity between a node and the previously seen nodes in the input sequence. Each node is thus supplied with an empirically fixed length adjacency vector of length $k$. Table~\ref{data} depicts these values of $k$ for each graph class.

To be able to produce layouts, the LSTM model is given the objective to optimize node coordinates with the help of a loss function. Starting with a 2D array of length $n$ with random $x$ and $y$ coordinates, at each iteration the model computes a specified loss. Through the use of gradient descent the model then attempts to minimize said loss, and changes the coordinates, by updating the network's weights. Equation \ref{PS} gives the Procrustes Statistic loss function.

\begin{tabular}{p{6cm}p{6cm}}
    \begin{equation}
    \label{PS}
    PS =  \sum_{i=0}^{n} (z_i - \hat{z_i})^2
    \end{equation}
    &
    \begin{equation}
    \label{SuS}
    SuS =  \frac{\sum_{k=0}^{n^2} (s_i - \hat{s_i})^2}{2}
    \end{equation}
\end{tabular}

With $Z = [z_0,...,z_n]$ and $\hat{Z} = [\hat{z_0},...,\hat{z_n}]$,
$Z$ and $\hat{Z}$ represent the transformed original and predicted node coordinates, respectively. The original and predicted node coordinates ($z_i$ and $\hat{z_i}$) are transformed through rotation, translation and scaling. In equation \ref{SuS}, that shows the Supervised stress loss function, $s_i$ and $\hat{s_i}$ denote the original and predicted stress for a pair of vertices, respectively, where $S = [s_0,...,s_{n^2}]$ and $\hat{S} = [\hat{s}_0,...,\hat{s}_{n^2}]$. Using matrix operations, $S$ can be computed as follows: $S = W * (E - D)^2$, where $D$ is the symmetric $nxn$ graph theoretical distance matrix, $W$ is equal to $D^{-2}$, and $E$ is the $nxn$ symmetric euclidean distance matrix.

\section{Graph classes}
\begin{figure}
\captionsetup{justification=raggedright,singlelinecheck=false}
    \centering
    \begin{subfigure}[t]{0.32\textwidth}
        \centering
        \includegraphics[width=\textwidth,trim={0 0 0 2cm},clip]{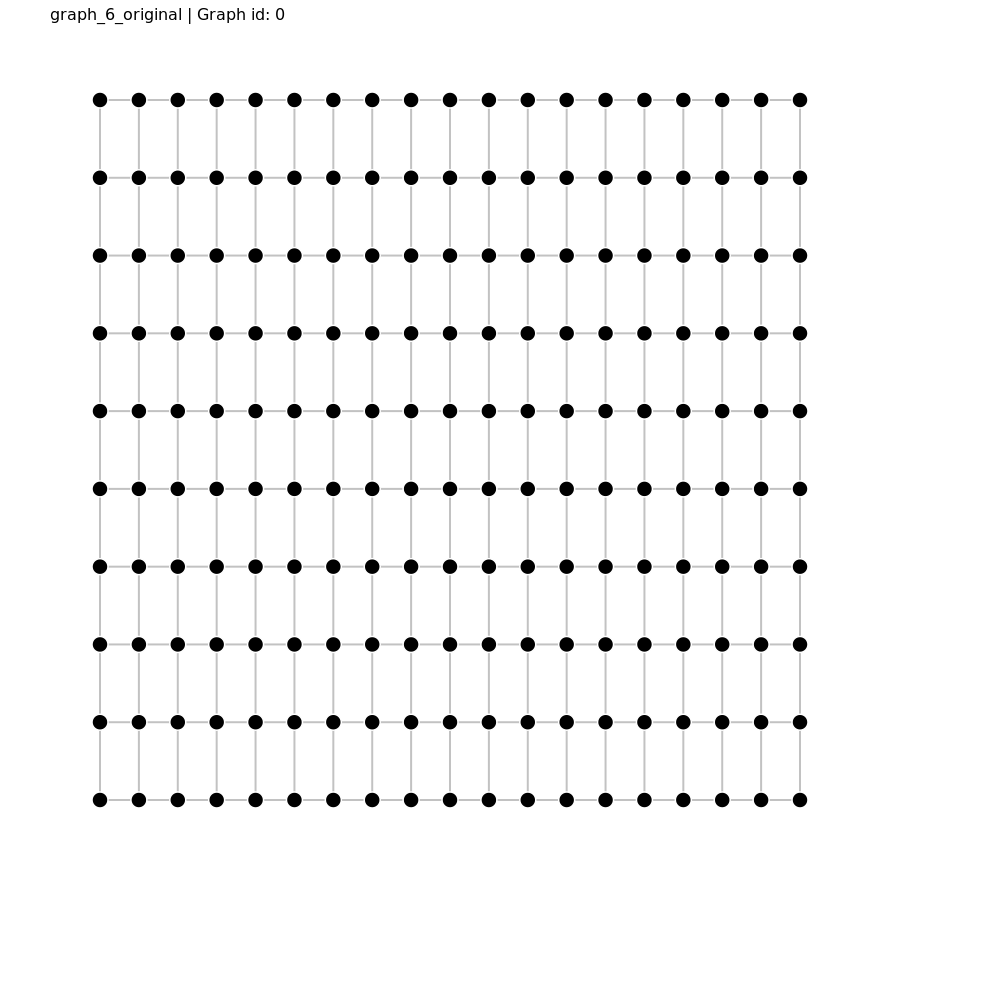}
        \caption{Grid graph}
        \label{fig:grid_example}
    \end{subfigure}
    \hfill
    \begin{subfigure}[t]{0.32\textwidth}
        \centering
        \includegraphics[width=\textwidth,trim={0 0 0 2cm},clip]{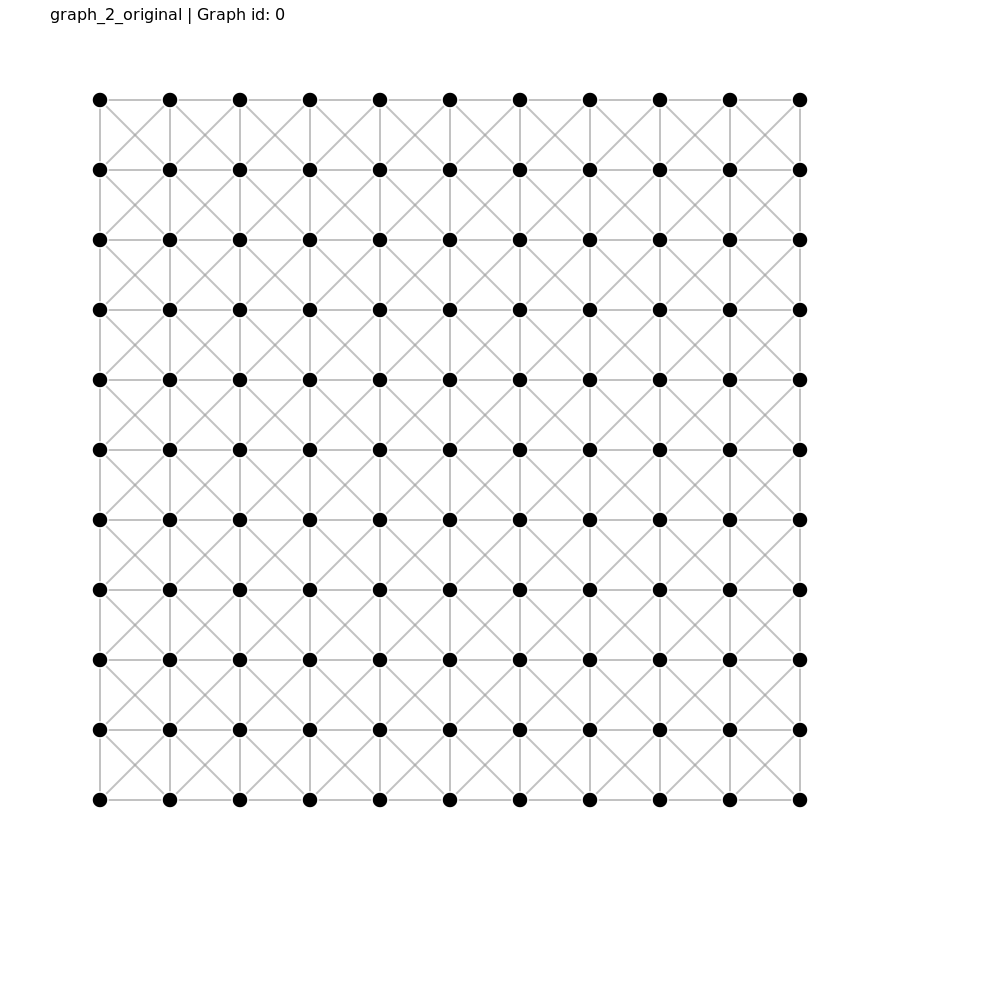}
        \caption{Grid graph with full diagonals ($Grids_d$)}
        \label{fig:grid_fulldiags_example}
    \end{subfigure}
    \hfill
    \begin{subfigure}[t]{0.32\textwidth}
        \centering
        \includegraphics[width=\textwidth,trim={0 0 0 2cm},clip]{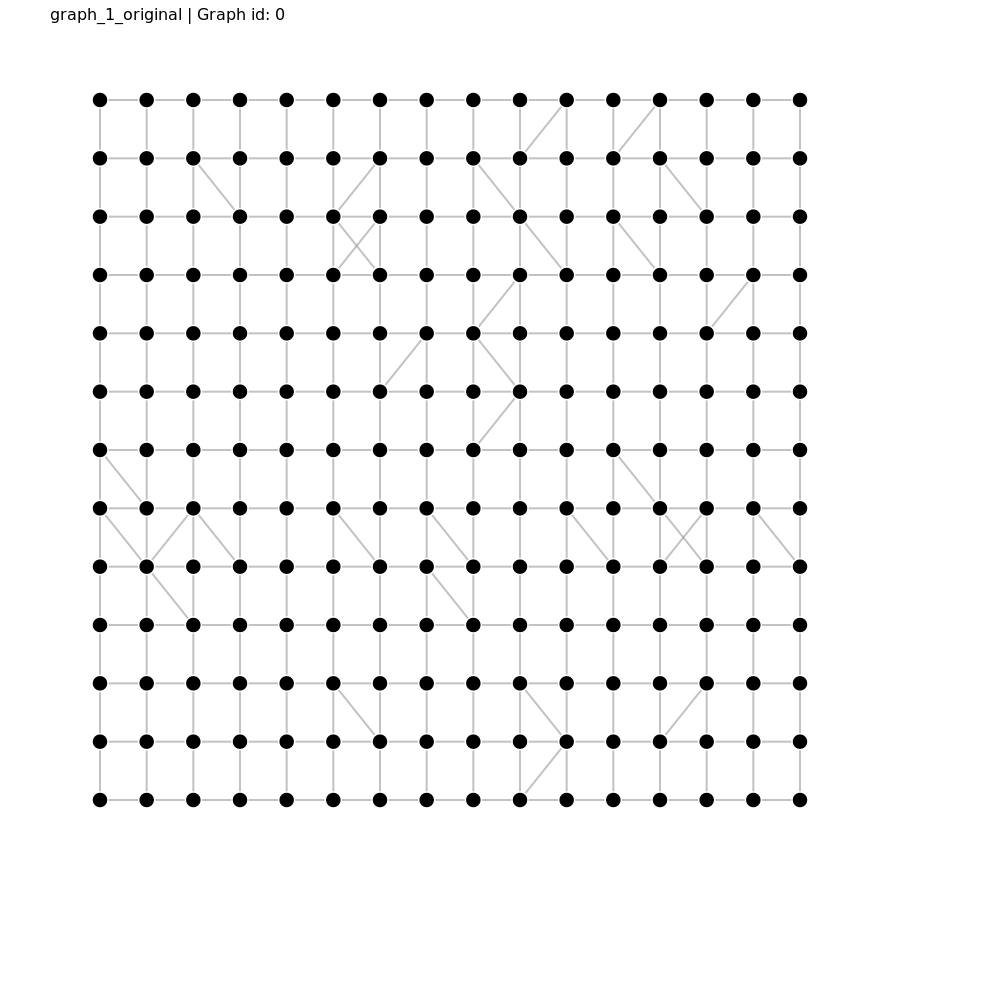}
        \caption{Grid graph with random diagonals ($Grids_{rd}$)}
        \label{fig:grid_randomdiags_example}
    \end{subfigure}
        \caption{Examples of grid type graphs}
        \label{fig:grid_graphs}
\end{figure}

\begin{figure}
    \centering
    \begin{subfigure}[t]{0.24\textwidth}
        \centering
        \includegraphics[width=\textwidth,trim={0 0 0 2cm},clip]{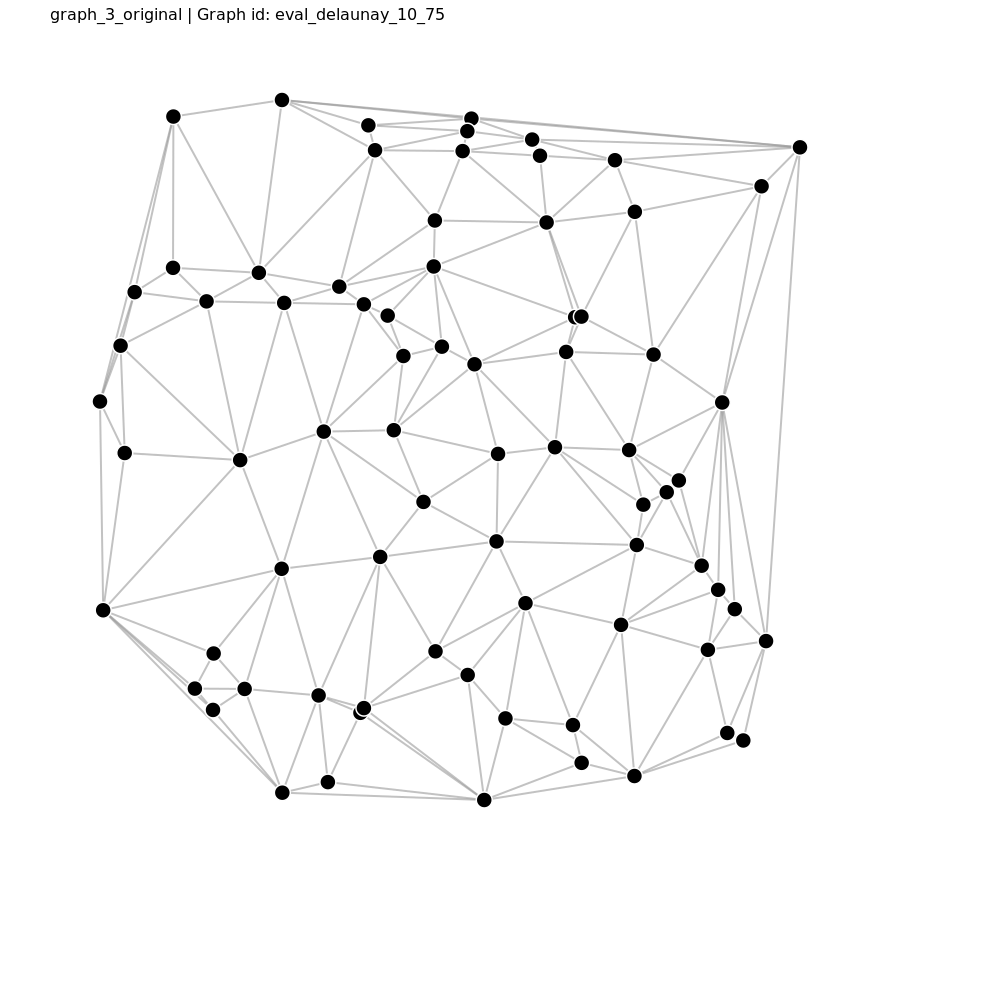}
        \caption{Delaunay triangulation}
        \label{fig:del_example}
    \end{subfigure}
    \hfill
    \begin{subfigure}[t]{0.24\textwidth}
        \centering
        \includegraphics[width=\textwidth,trim={0 0 0 2cm},clip]{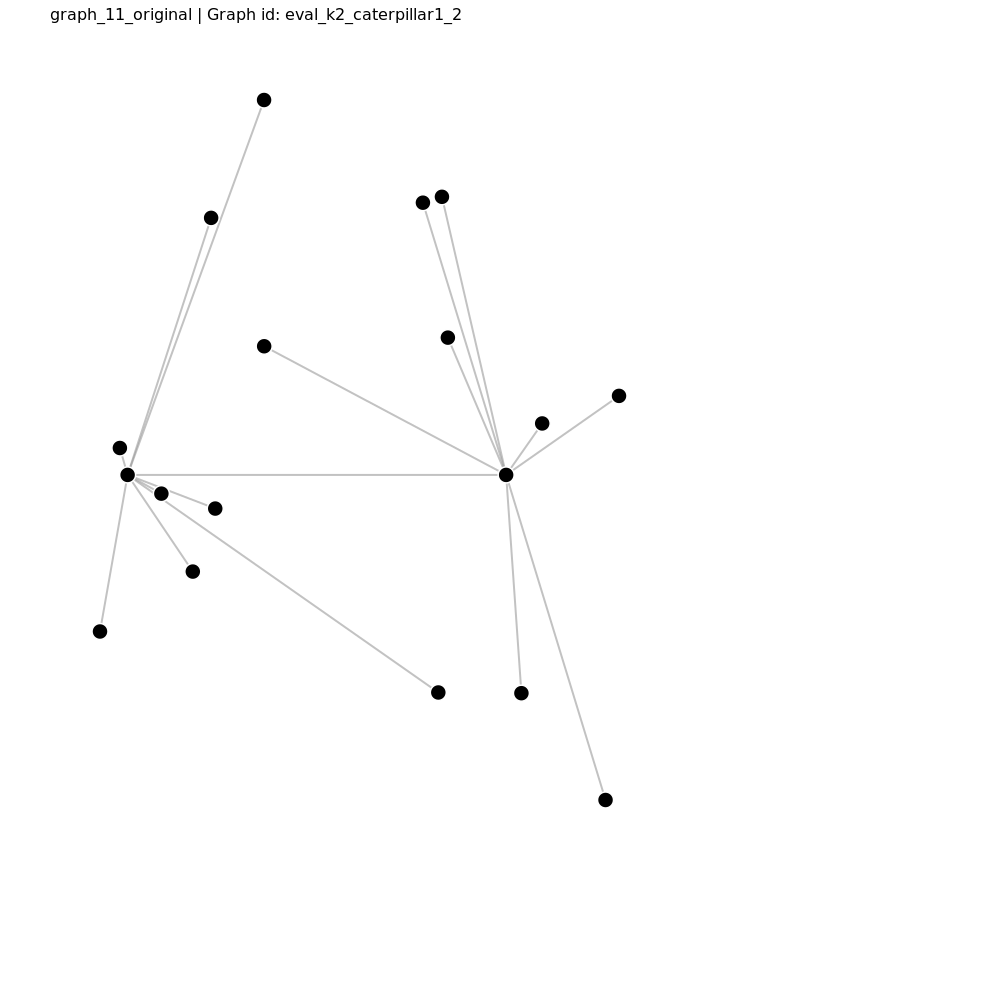}
        \caption{2-star caterpillar}
        \label{fig:2cat_example}
    \end{subfigure}
    \hfill
    \begin{subfigure}[t]{0.24\textwidth}
        \centering
        \includegraphics[width=\textwidth,trim={0 0 0 2cm},clip]{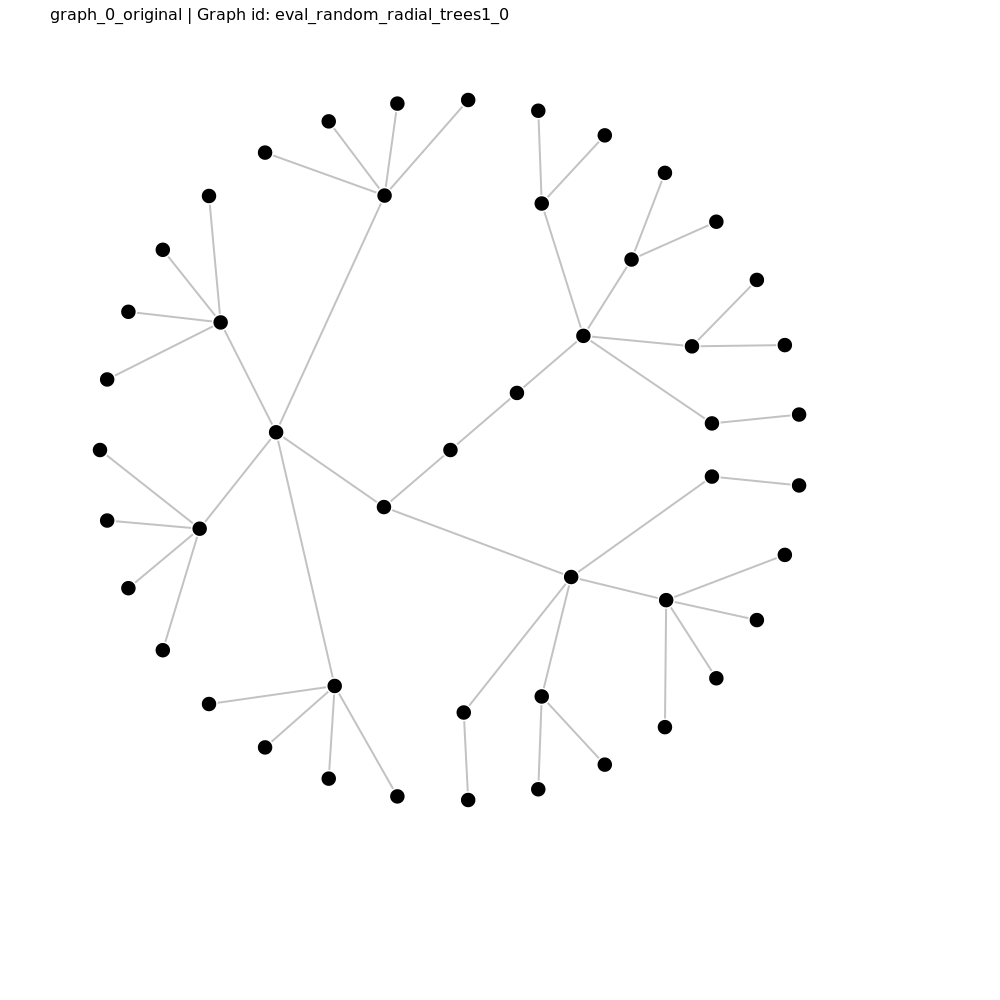}
        \caption{Randomized radial tree}
        \label{fig:rrtree_example}
    \end{subfigure}
    \hfill
    \begin{subfigure}[t]{0.24\textwidth}
        \centering
        \includegraphics[width=\textwidth,trim={0 0 0 2cm},clip]{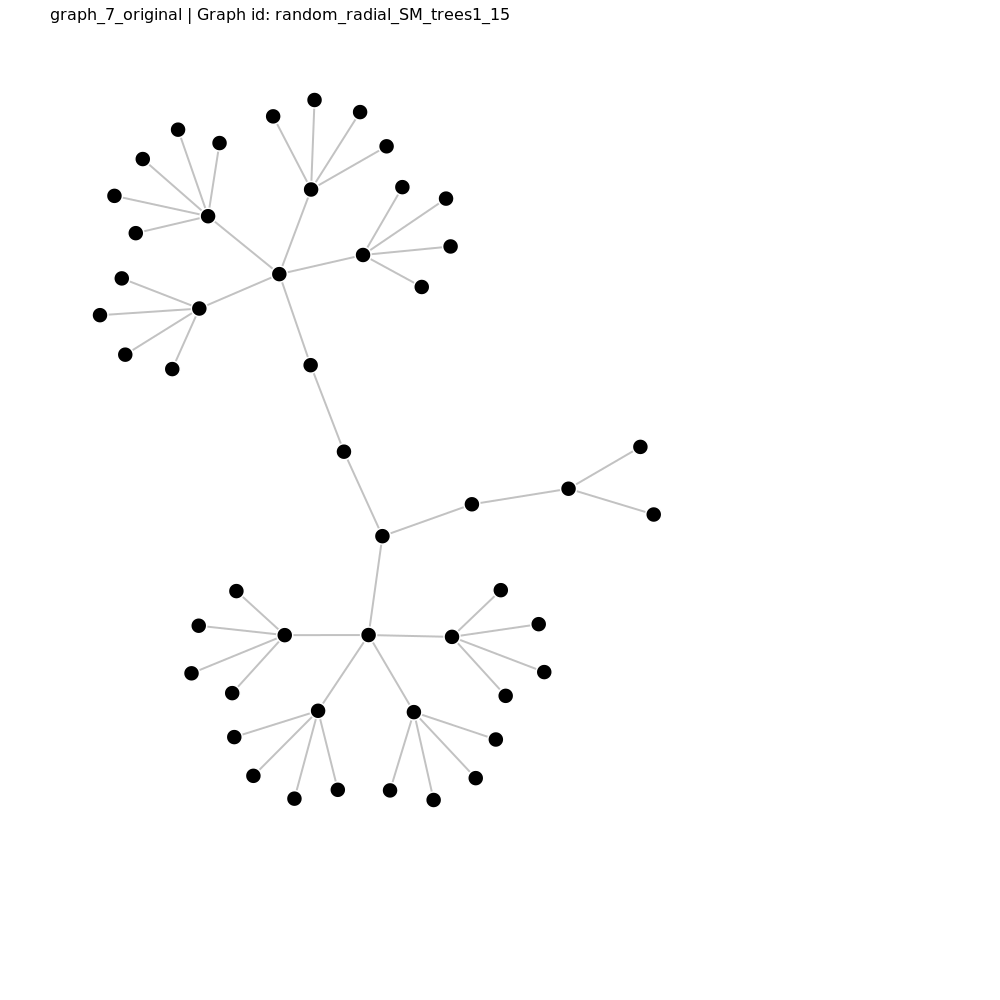}
        \caption{Randomized tree with SM}
        \label{fig:rrsmtree_example}
    \end{subfigure}
        \caption{Examples of other graph classes}
        \label{fig:rest_graphs}
\end{figure}

In our experiments we have applied the model to the following graph classes.
\paragraph{Grids}
Rectangular $n \times m$ grids. Directly taken from Wang et al. \cite{deepdrawing,deepdrawing_github}. See Figure~\ref{fig:grid_example}.
\paragraph{Grids$_d$}
Similar to regular grids except every cycle of length 4 has two additional edges connecting its opposite vertices, referred to as 
\emph{diagonals} in the paper. See Figure~\ref{fig:grid_fulldiags_example}.
\paragraph{Grids$_{rd}$}
Similar to regular grids except that each diagonal is created with probability $p = 0.05$. See Figure~\ref{fig:grid_randomdiags_example}.
\paragraph{Delaunay}
Nodes are represented by $n$ random points on the plane and edges by applying a Delaunay triangulation algorithm. See Figure~\ref{fig:del_example}.
\paragraph{Caterp}
A $k$ number of stars are created and linked to each other via a single edge. The star centers are fixed and a random number of tips of each star are given random angles and a random length within a certain lower and upper bound ($lb,ub$). See Figure~\ref{fig:2cat_example}.
\paragraph{RR Trees}
Starting with 1 parent node and a default number of children of $c = 4$, with a probability of $p = 0.10$ a random number of children within the boundary ($2, ub$) are connected to the parent node. This process is repeated for the children until $l$ levels are completed. The resulting graph's layout is produced with a radial layout algorithm~\cite{radial_layout}. See Figure~\ref{fig:rrtree_example}.
\paragraph{RSM Trees}
Similar to RR Trees but the resulting graph's layout is produced with a Stress Majorization layout algorithm. See Figure~\ref{fig:rrsmtree_example}.

\section{Experimental setup \& results}
\subsection{Setup}
\subsubsection{Data}
We train each model on a separate training data set with one loss function. With 8 different graph classes and 2 loss functions, we therefore end up with 16 different models. Table \ref{data} displays an overview of the training and testing datasets of each graph class.

\begin{table}[H]
\caption{Test and train datasets statistics\label{data}}
\begin{tabular}{||l||l|l|l|l|l|l|l|l||}
\hline
Graph class & \#Nodes & \#Edges & $k$ & Train & Validation & Test & \#Instances & \#Nodes in instances\\
\hline
\hline
Grids & [100, 576] & [180, 1104] & 49 & 72 & 24 & 24 & NA & NA\\
Grids$_d$ & [100, 576] & [180, 1680] & 49 &72 & 24 & 24 & NA & NA\\
Grids$_{rd}$ & [100, 576] & [180, 1242] & 49 &72 & 24 & 200 & $50 \times 4$ & [120, 208, 240, 360]\\
Delaunay & [25, 100] & [70, 282] & 30 & 1000 & 200 & 200 & $50 \times 4$ & [25, 50, 75, 100]\\
Caterp2 & [18, 116] & [17, 115] & 113 & 400 & 50 & 200 & $50 \times 4$ & [18, 38, 58, 78]\\
Caterp3 & [26, 176] & [25, 175] & 113 & 400 & 50 & 200 & $50 \times 4$ & [26, 44, 62, 80]\\
RR Trees & [50, 625] & [49, 624] & 397 & 500 & 50 & 200 & $50 \times 4$ & [50, 100, 150, 200]\\
RSM Trees & [50, 625] & [49, 624] & 397 & 500 & 50 & 200 & $50 \times 4$ & [50, 100, 150, 200]\\
\hline
\end{tabular}
\end{table}

Most testing datasets are comprised of multiple instances of a graph of the same size. E.g.: The testing dataset of RR Trees contained 50 instances of a graph with 50 nodes, 50 instances of a graph with 100 nodes etc. Having multiple instances of the same node size allows for more valid comparisons with the FD and SM layout techniques.

\subsubsection{Implementation \& configuration}
The PyTorch Python implementation of Wang et al.~\cite{deepdrawing_github} is reused. The LSTM model is run on a machine with a Quadro RTX 3000 GPU and an Intel Core i7-10850H CPU (2.7GHz). The learning rate and batch size of each model was set to 0.0015 and 24, respectively.

\subsection{Results}
\subsubsection{Model results}

Table \ref{fulltab_results} contains the results of the model trained on different graph classes using the Procrustes Statistic loss function.

\begin{table}[H]
\caption{Averaged performance of conventional techniques and model with the Procrustes Statistic LF on multiple instances of different graph classes. Bolded entries indicate interesting differences.\label{fulltab_results}}
\begin{tabular}{||l|l||l|l|l|l||l|l|l|l||l|l|l|l||}
\hline
Graph class & & LF & QM & QM & QM & & QM & QM & QM & & QM & QM & QM\\
\hline
Train & Test & PS & nc & s & ar & FD & nc & s & ar & SM & nc & s & ar\\
\hline
\hline
Grids & Grids & 0.010 & \textbf{36} & 6.91 & 0.051 &  & \textbf{15.90} & 6.18 & 0.50 & & \textbf{0.61} & 4.88 & 0.93\\
Grids\_d & Grids\_d & 0.017 & \textbf{352} & 13.60 & 0.064 & & \textbf{417} & 12.40 & 0.19 & & \textbf{426} & 9.55 & 0.73\\
Grids\_rd & Grids\_rd & 0.0038 & 6.74 & 7.06 & 0.072 & & 27.5 & 6.10 & 0.26 & & 3.04 & 4.82 & 0.57\\
Grids & Grids\_rd & 0.043 & 385 & 6.32 & \textbf{0.0033} & & 27.5 & 6.10 & \textbf{0.26} & & 3.04 & 4.82 & \textbf{0.57}\\
Grids\_d & Grids\_rd & 0.23 & 858 & 5.37 & 0.0019 & & 27.5 & 6.10 & 0.26 & & 3.04 & 4.82 & 0.57\\
\hline
Delaunay & Delaunay & 0.30 & 236 & 3.35 & 0.0081 & & 59.40 & 3.57 & 0.026 & & 90 & 3.10 & 0.038\\
\hline
Caterp2 & Caterp2 & 0.60 & 0 & 5.04 & 0.021 & & 0 & 4.24 & 0.090 & & 0 & 5.39 & 0.15\\
Caterp3 & Caterp3 & 0.52 & 0.16 & 2.86 & \textbf{0.0028} & & 0.39 & 3.75 & \textbf{0.061} & & 0 & 3.98 & \textbf{0.067}\\
\hline
RR Trees & RR Trees & 0.53 & 176 & 5.53 & 0.0047 & & 3.87 & 5.13 & 0.052 & & 30.90 & 5.63 & 0.13\\
RSM Trees & RSM Trees & 0.49 & 229 & 4.96 & 0.0031 & & 3.68 & 5.28 & 0.048 & & 32.80 & 5.66 & 0.13\\

\hline
\end{tabular}
\end{table}

\end{subappendices}

\end{document}